\documentclass[a4paper,fleqn,usenatbib,useAMS]{mnras}

\usepackage{newtxtext,newtxmath}

\usepackage[T1]{fontenc}

\usepackage{enumitem}
\usepackage{mathtools}
\usepackage{graphicx}
\usepackage{amssymb}
\usepackage{amsmath}
\usepackage{ulem}
\usepackage{epstopdf}
\usepackage{graphicx}
\usepackage{xcolor}

\def\lambar{\lambda\llap {--}}

\title[Ultra long period magnetars]{Periodicity in recurrent fast radio bursts and the origin of ultra long period magnetars}

\author[P. Beniamini et al.]{
Paz Beniamini$^{1}$\thanks{E-mail: paz.beniamini@gmail.com},
Zorawar Wadiasingh$^{2,3,4}$,
Brian D. Metzger$^{5,6}$\\
$^{1}$Division of Physics, Mathematics and Astronomy, California Institute of Technology, Pasadena, CA 91125, USA\\	
$^{2}$Astrophysics Science Division, NASA Goddard Space Flight Center, Greenbelt, Maryland, 20771, USA\\
$^{3}$Universities Space Research Association (USRA) Columbia, Maryland 21046, USA\\
$^{4}$Centre for Space Research, North-West University, Potchefstroom, South Africa\\
$^{5}$Department of Physics and Columbia Astrophysics Laboratory, Columbia University, New York, NY 10027, USA\\
$^{6}$Center for Computational Astrophysics, Flatiron Institute, New York, NY 10010, USA}

\begin{document}
\label{firstpage}
\pagerange{\pageref{firstpage}--\pageref{lastpage}}
\maketitle

\begin{abstract}
The recurrent fast radio burst FRB 180916 was recently shown to exhibit a 16 day period (with possible aliasing) in its bursting activity.  Given magnetars as widely considered FRB sources, this period has been attributed to precession of the magnetar spin axis or the orbit of a binary companion.  Here, we make the simpler connection to a {\it rotational period}, an idea observationally motivated by the 6.7 hour period of the Galactic magnetar candidate, 1E 161348--5055. We explore three physical mechanisms that could lead to the creation of ultra long period magnetars: (i) enhanced spin-down due to episodic mass-loaded charged particle winds (e.g. as may accompany giant flares), (ii) angular momentum kicks from giant flares and (iii) fallback leading to long lasting accretion disks. We show that particle winds and fallback accretion can potentially lead to a sub-set of the magnetar population with ultra long periods, sufficiently long to accommodate FRB 180916 or 1E 161348--5055. If confirmed, such periods implicate magnetars in relatively mature states (ages $1-10$ kyr) and which possessed large internal magnetic fields at birth $B_{\rm int}\gtrsim 10^{16}$ G. In the low-twist magnetar model for FRBs, such long period magnetars may dominate FRB production for repeaters at lower isotropic-equivalent energies and broaden the energy distribution beyond that expected for a canonical population of magnetars which terminate their magnetic activity at shorter periods $P \lesssim 10$ s.
\end{abstract}

\begin{keywords}
fast radio bursts -- stars: magnetars -- stars: winds, outflows -- stars: magnetic field -- accretion, accretion discs
\end{keywords}


\defcitealias{Wadiasingh2020}{W20}
\section{Introduction}
\label{sec:Intro}
Fast radio bursts (FRBs) are short ($\sim$ ms long) radio signals whose origin and production mechanism, though the source of much speculation, are yet to be understood \citep{Lorimer+07,Thornton+13,2018arXiv181005836P,2019A&ARv..27....4P}. Starting with FRB~121102 \citep{Spitler+14}, several FRBs have been observed to repeat \citep{CHIME+19REPEATERS}. Recently, the Canadian Hydrogen Intensity Mapping Experiment (CHIME) experiment reported the discovery of a $\sim 16$-day periodicity in the bursts from one such repeater, FRB 180916 \citep{2020arXiv200110275T}.
Within this apparent periodicity, there is a 4-day `active phase' window within which all bursts are detected.
Owing to the regular intermittent exposure of CHIME, it is not possible from the current observations to exclude higher frequency aliases of the 16 day period\footnote{Allowed periods range from one hour to the full 16 days, and with a slight statistical preference towards the 16 day period for the true periodicity.}.  Nonetheless, this is the first\footnote{\cite{2020arXiv200303596R} also report a tentative periodicity of $\sim 160$ days in FRB 121102 with a much wider active window. However, this result is less secure than the CHIME periodicity and will require a longer baseline to confirm or refute.} periodic signal seen in an FRB and may provide crucial information for deciphering the FRB mystery.  

Thus far, two main hypotheses have been raised to explain the periodicity, both involving a highly magnetized neutron star, or "magnetar", as the FRB source. In the first, the magnetar is in a tight binary with an early OB-type star that has a strong wind which obscures the FRB radiation, except through a rather narrow channel \citep{2020arXiv200201920L,Ioka2020}. In the second, the magnetar is undergoing free precession \citep{2020arXiv200205752Z,2020arXiv200204595L} due to a slight non-spherical deformation of the magnetar.  A third (and perhaps the simplest) possibility, that the periodicity represents the rotation period of the magnetar, has been dismissed, due to the fact that prominent Galactic magnetars possess significantly shorter spin periods, $P \lesssim 12$ s. Of course, a similar argument can be applied to the other two scenarios, as it is also the case that none of the Galactic magnetars reside in binaries or exhibit free precession. Regardless, this conclusion is precarious, for three reasons we discuss below.

First, at least one candidate magnetar, the central compact object in the supernova remnant RCW 103, possesses an astonishingly large period of 6.67 hr \citep{2006Sci...313..814D}. This putative magnetar, 1E 161348--5055 possesses most features of the more rapidly-spinning Galactic magnetars\footnote{\cite{2011MNRAS.418..170E} report $\dot{P} < 1.6 \times 10^{-9}$~s s${}^{-1}$.}, including millisecond duration short bursts \citep{2016MNRAS.463.2394D}, longer-term outbursts, and non-thermal (and relatively flat) broad-pulsed power-law X-ray emission beyond $20$ keV with {\it{NuSTAR}}  \citep{2016ApJ...828L..13R} characteristic of strong-field resonant Compton scattering \citep{2007Ap&SS.308..109B,2007ApJ...660..615F,2018ApJ...854...98W}. {\it{Chandra}} imaging reveals a proper motion of $169 \pm 51$ km s${}^{-1}$ \citep{2017ApJ...844...84H}, suggesting the neutron star received a sufficiently large natal kick to disrupt any wide binary during its birthing supernova. Indeed, limits from the {\it Hubble Space Telescope} exclude a companion star hotter than M7 ($T_{\rm eff} \gtrsim 2800$ K), ruling out an accretion-powered scenario for the emission from 1E 161348--5055 \citep{2017ApJ...841...11T}. Finally, it is worth noting that other magnetar candidates (discovered via their bursts) exist with unconstrained $P$, which could in principle be as slowly spinning as 1E 161348--5055.

Second, even if ultra long period (ULP) magnetars with $P \gg 10$ s are rare among the population of extragalactic magnetars, some emission models suggest they will be particularly efficient FRB producers and as such, significantly over-represented in the FRB population (this is the situation for example in the low-twist model as will be shown later in this paper). Several weeks after the submission of this paper, the first Galactic FRB, FRB 200428, has been detected \citep{2020ATel13684....1B}, arising from the Galactic magnetar SGR 1935+2154. SGR 1935+2154 has a period of $3.2$ s, typical for Galactic magnetars. Interestingly, the repetition rate of similar such bursts from SGR 1935+2154, can be strongly constrained to be $\sim 10^5$ times lower than that of FRB 180916 \citep{Margalit2020}. This is consistent with the scenario presented in this paper, suggesting that ULP magnetars could be much more prolific FRB sources than `standard' magnetars.

Third, additional support in favour of the rotation model for FRB periodicity comes once more from the observations of the recent Galactic FRB. These suggest that not all magnetar bursts with energies comparable or larger than the burst associated with FRB 200428 are accompanied by FRBs \citep{Margalit2020,Lu2020}, including the giant flare from SGR 1806-20 \citep{Tendulkar2016}. This implies that either (i) intrinsically some flares do not produce FRBs; or (ii) all flares produce FRBs but they are generally beamed away from us. Scenario (i) is constrained by the activity level of some of the most prolific FRBs, e.g. 121102 \citep{Law2017}, which (assuming a X-ray/radio energy ratio similar to that observed in FRB 200228) is already comparable to the rate of flares for a $\gtrsim 10^{16}$G field magnetar \citep{Margalit2020}. Instead, if (ii) is true, then the detection of a given FRB may be modulated by the rotational phase. By contrast, we know that Galactic magnetars neither have binary companions nor precession which suggests that the rotationally-driven detection picture is more consistent with the population of Galactic magnetars than the other scenarios for the 16-day periodicity.

Beyond their potential explanation for periodicity in FRBs, the origin of ULP magnetars such as 1E 161348--5055 is of interest in its own right. In this paper, we explore three potential formation channels.  The first involves a mass-loaded charged particle wind from the magnetar, which when active, e.g. in the aftermath of giant flares (GF) expands the open magnetic flux of the star and thus temporarily enhances its spin-down rate over that of ordinary dipole spin-down. The second channel is a cumulative loss of rotation due to angular momentum kicks imparted through asymmetric GF emission. The third channel involves late-time fallback of high angular momentum matter into a disk surrounding the magnetar from its birthing supernova. As we will show, one or a combination of these scenarios could account for modest fraction of ULP magnetars. 

\section{Implications of different interpretations of FRB periodicity}

We assume that FRBs arise from flare-like events on magnetars, as is supported by their non-Poissonian arrival times and power-law fluence distributions of the bursts from the well-studied repeater FRB 121102\footnote{See, for example, the next-burst inter-arrival time distribution for the 2017 burst storm of FRB 121102 detected by GBT \citep{2018ApJ...866..149Z}, plotted in figure~2 of \citet[][]{2019ApJ...879....4W}, versus the equivalent for Galactic magnetar 1E 2259+586 displayed in figure 10 of \citet{2004ApJ...607..959G}.}. This emission may originate directly from the magnetosphere (e.g.~\citealt{Kumar2017,LuKumar2018,2019ApJ...879....4W}), or at much larger radii from shocks generated as flare ejecta collides with the particle wind surrounding the magnetar (\citealt{2014MNRAS.442L...9L,Beloborodov17,Metzger+19}). Aside from its lower energetics, the pulses of FRB 180916 and their respective fluence distributions are otherwise similar to FRB 121102 and other CHIME repeaters, suggesting similar progenitors.

In this work, we suggest that the observed periodicity arises due to the spin of the magnetar. The requisite beaming of the FRB emission is then provided either by the finite width of pulsar-like polar cap beaming (as in regular pulsars); or due to a preferential direction for magnetic flare ejecta relative e.g.~to the magnetic dipole axis (in maser shock scenarios). The fact that no apparent periodicity is seen in FRB 121102 \citep{2018ApJ...866..149Z} over a contiguous observation of $\sim 5$ hours suggests either a wide beaming cone for the FRB emission or long periodicity $>5$ hours. There also appears to be frequency selection in FRB 180916 -- two bursts seen by CHIME between 400--800 MHz were not observed in simultaneous observations by Effelsberg at 1234--1484 MHz \citep{2020arXiv200110275T}. Such frequency filtering is consistent with a neutron star magnetospheric scenario as in canonical pulsar radius-to-frequency mapping \citep[e.g.,][]{1970Natur.225..612K,1978ApJ...222.1006C}, and would be rotational-phase dependent. If frequency filtering is indeed important in recurrent FRBs (i.e. if the lack of high frequency bursts is due to the rotation of the magnetosphere and not due to the emission process) it would support beamed pulsar-like emission, which in turn, would imprint periodicity in the FRB emission on the spin period.  In this picture, if FRB 121102 is similar to FRB 180916, the absence of shorter periodicity suggest that FRB 121102 is not a magnetar with canonical $P\sim 1-10$ s.

\subsection{Shrouded Binary Scenario}
\cite{2020arXiv200201920L,Ioka2020} propose a shrouded binary magnetar scenario for FRB 180916, in which the periodicity in FRB arrival-times arises due to eclipses by a massive star wind which partially engulfs the magnetar (which in their scenario possesses a canonical spin period $P \sim 1-10$ s). The situation is similar to the eclipses of PSR B1259--63 and analogous to ``redback" millisecond pulsar binaries where the intrabinary shock envelops the pulsar \cite[e.g.,][]{2017ApJ...839...80W,2018ApJ...869..120W}. 

For an eclipsing/shrouding scenario, a greater number of high frequency bursts should be seen (unless their intrinsic production rate or detectability is lower) because the absorption/scattering opacity of radio waves in plasma generally decreases with increasing observing frequency $\nu_{\rm obs}$\footnote{Study of the eclipsing mechanisms for high brightness temperature pulsar-like emission is not a straightforward exercise, and may involve several wave-particle plasma processes \citep[e.g.,][]{1991ApJ...370L..27E,1994ApJ...422..304T}. }. Indeed, in known eclipsing pulsar systems, the eclipse duration relative to the orbital period scales as $f_E \propto \nu_{\rm obs}^{-0.4}$ \citep{1990ApJ...351..642F,2016MNRAS.459.2681B,2018MNRAS.476.1968P,2020arXiv200302335P} so that the window of the uneclipsed phase is longer at higher $\nu_{\rm obs}$. 
So far however there is a paucity of Effelsberg bursts in phases where the majority of CHIME bursts are observed, in contrast to the expectation above.  

The orbital phase should affect also the flux of observed bursts. The center of the uneclipsed window
is where flux density is expected to be highest \citep[e.g.][]{2016MNRAS.459.2681B} in the binary scenario. However, there is no significant variation of fluence across the 4-day window, suggesting a much sharper transition than shrouding by a diffuse and turbulent two-wind interaction. Moreover, the dispersion measure (DM) of the Effelsberg bursts at the edges of the 4-day window (i.e. eclipse periphery) are similar to that of the CHIME bursts, suggesting a variation $\Delta$DM $\lesssim 0.1$ pc cm$^{-3}$ over the observed window. Yet, in known millisecond pulsar eclipsing systems (which are more compact with lower mass companions and therefore presumably lower plasma columns), the eclipses often exceed this value before the source flux density degenerates into the noise level \citep{1991ApJ...380..557R,2001MNRAS.321..576S,2009Sci...324.1411A,2013arXiv1311.5161A,2018JPhCS.956a2004M}. 
Besides, in PSR B1259--63, the closest known analog to a putative magnetar-massive star binary, the DM variation may exceed $\Delta$DM $\gtrsim 10$ pc cm$^{-3}$ while typically $\Delta$DM $\sim 1-5$ \citep{2005MNRAS.358.1069J} \footnote{Although the orbital period of PSR B1259--63 is significantly longer than $16$ days, its orbit is highly eccentric and the eclipsing orbital phase in PSR B1259--63 is only $\sim 30$ days near periastron.}. 

\subsection{Precession Scenario}
An alternative source of periodicity can be due to free precession of the magnetar \citep{2020arXiv200205752Z,2020arXiv200204595L}. A large internal field $B_{\rm int} \sim 10^{16}$ G and high core $T\gtrsim 10^9$ K are required to achieve a large ellipticity. This in turn requires a young magnetar.

Precession models {\it necessarily} invoke beaming along a preferred direction (e.g. magnetic axis) for the FRB emission process. A canonical magnetar with $P\sim 1-10$ s, will then exhibit a shorter scale periodicity (with period $P$) within each 4-day precession period window, given sufficient statistics. The required statistics to resolve this signature are not yet available for FRB 180916.
However, the precession model cannot account for the lack of shorter periodicity \citep{2018ApJ...866..149Z} (due to the spin) in the more prolific bursts of FRB 121102 without either abandoning the assumption of beaming or ascribing FRB 121102 to a different class of progenitor.

Since the precession period scales inversely with the magnetar's ellipticity, if the model is correct, many more FRBs with longer apparent periodicities ought to be observed. Indeed, if the true period of FRB 180916 is significantly lower than 16 days, due to frequency aliasing (see \S \ref{sec:Intro}), it will become much more difficult to explain the periodicity with precession models. Finally, in the precession model, the normal polarization angle should sweep at a period of $P$ along with a slow secular change in magnetic obliquity at the precession period \citep{2020arXiv200205752Z}.
 

 \section{Enhanced spin-down following giant flares}
  \subsection{Observational evidence}
  \label{sec:Directobs}
  Direct empirical evidence supporting enhanced spin-down in GFs was the observed spin frequency decrease $\Delta \Omega/\Omega\sim - 10^{-4}$ following the GF of SGR 1900+14, which released an energy in gamma-rays of $\sim 4\times 10^{44}$ erg \citep{Tanaka2007}. \cite{Thompson2000} discuss the possibility that this spin-period increase is due to the mass-loaded charged particle wind (see \S \ref{sec:motivatePwind}), as well as an alternative possibility (which they moderately favor) that the spin change is due to angular momentum exchange between the crust and the rest of the magnetar. The most energetic GF seen to date was observed in 2004 from SGR 1806-20 \citep{2005Natur.434.1107P}. Over almost 30 years of observations, SGR 1806-20 has been shown to spin-down at a faster rate than that extrapolated from its historical evolution \citep{Younes2015}. By 2012, its spin frequency has decreased by $\sim 2\%$ compared to the extrapolated rate from 1994.  Another example, is the magnetar 1E 2259+586, which exhibited an `anti-glitch' with the spin frequency decreasing as $\Delta \Omega/\Omega=-10^{-6}$ over a timescale of $\lesssim 100$ days \citep{Archibald2013}.  This weaker spin frequency decrease was not concurrent with a GF, but could be an indication that the mass-loaded winds discussed below (at lower luminosities than associated with flares) are ubiquitous in magnetars, even if their physical origin remains opaque. More recently, Younes et al. (in prep) detected a similar spin-down glitch in 1E 2259+586 with NICER without any large radiative changes.
 
 Magnetar flares feed off the magnetar's reservoir of magnetic energy $E_{\rm B} \sim (4\pi R_{\rm NS}^{3}/3)(B_{\rm int}^{2}/8\pi)$, where $B_{\rm int}$ is the average internal magnetic field strength (which can be larger than the energy contained in the external dipole field). The total number of flares can be crudely estimated (we adopt a more realistic flare energy distribution in \S \ref{sec:MonteCarlopw}) as $N=E_{\rm B}/E_{\rm f}$. Taking the above empirically-measured frequency changes at face value, and assuming a constant fractional frequency decrease $f_{E_{\rm f}} =-\Delta \Omega/  \Omega_0$ accompanies all flares of energy $E_{\rm f}$ (this is the expectation in the mass-loaded wind scenario, as shown in equation \ref{eq:Pf} below), the present-day frequency is estimated as, $\Omega_f=\Omega_0\exp(-f_{E_{\rm f}} E_{\rm B}/E_{\rm f})$.  Thus, if $E_{\rm B}\gtrsim f_{E_{\rm f}}^{-1} E_{\rm f}$ the final spin period may be significantly enhanced due to its evolution during, or following, GFs. 
 
 As an illustration, consider an internal magnetic field of $B_{\rm int}=10^{16}$ G, leading to $E_{\rm B}=3\times 10^{49}\mbox{ erg}$. This energy can be dissipated by $\sim 10^5$ flares with $E_{\rm f}\sim 3\times 10^{44}\mbox{ erg}$ comparable to the GF of SGR 1900+14\footnote{Most magnetars are likely to be born with significantly lower $B_{\rm int}$ and would be able to experience much less such GFs, scaling as $B_{\rm int}^2$ during their lifetime before depleting the magnetic energy}. Assuming that each one corresponds to the same fractional frequency change (see equation \ref{eq:Pf}) of $\Delta\Omega/\Omega=-10^{-4}$, leads to a decrease of $\Omega$ by a factor of $\sim 2\times 10^3-10^4$ as compared to the initial spin frequency, demonstrating that it is plausible for magnetars to rid themselves of the vast majority of their angular momentum due to spin-down following GFs. We return to this point in more quantitative detail in \S \ref{sec:MonteCarlopw}.
 If the absolute magnitude of the spin decrease (rather than the fractional spin decrease) is constant between GFs (as is the expectation in the case of spin-down due to kicks, discussed in \S \ref{sec:kicks} below), then the spin-frequency would decrease even faster (and in fact vanish within a finite time) than suggested by the exponential relation above. 

 \subsection{Spin-down due to kicks}
 \label{sec:kicks}
One mechanism that could cause spin-down is asymmetric energy ejection during GFs. We briefly demonstrate below that even with modest asymmetry, large amounts of angular momentum can be removed from a magnetar during each GF.
 
Consider a flare of energy $E_{\rm f}$, ejected from a point separated by some distance $r=fR_{\rm NS}$ from the spin axis. Here $f$ is some dimensionless number that encompasses the asymmetry of the flare ejection ($f=0$ corresponds to a fully symmetric ejection). For $f>0$ the GF takes with it angular momentum from the magnetar as it is emitted.  Up to a geometric factor of order unity, and assuming the duration of the flare, $t_{\rm f}\sim 0.2$ s, is close to instantaneous as compared to the magnetar's period (this is true during most of the magnetar's life, once its initial spin has sufficiently decayed due to dipole radiation), the change in spin-frequency due to this process is
\begin{equation}
     |\Delta \Omega|=\frac{fR_{\rm NS}E_{\rm f}}{cI}=2.5\times 10^{-5}E_{\rm f,45}f\mbox{ s}^{-1}
\end{equation}
where $I$ is the moment of inertia of the magnetar and where unless otherwise specified we adopt here and elsewhere the notation $x  \equiv x_y  \, 10^{y}$ in cgs units. We thus find that even if the asymmetry of the flare ejection is on a rather small scale (of the order of the neutron star radius), it is possible to cause a reduction of the spin-frequency that is comparable to the anti-glitch seen in SGR 1900+14 discussed in \S \ref{sec:Directobs} above. However, since the observed anti-glitch in SGR 1900+14 occurred on a timescale significantly longer than the duration of the GF itself, we slightly disfavour this scenario for that period change. 
In order for the spin-change from consecutive flares to add up coherently, rather than through random walk (which would require $(\Delta \Omega_{\rm tot}/\Delta \Omega)^2$ rather than the much lower $\Delta \Omega_{\rm tot}/\Delta \Omega$ kicks to accumulate a given change in spin of $\Delta \Omega_{\rm tot}$), some level of asymmetry is required in the direction of the ejected flare relative to the spin axis. This could happen if, for example, the magnetar is an oblique rotator and the GF is preferentially produced along the magnetic axis or alternatively if there is a crustal defect that creates a preferred point on the magnetar's surface from which GFs are released.

The change in the magnetar's spin would also be followed by a change in linear momentum. In fact, the latter, does not require any degree of asymmetry in the flare ejection location, and is given by $v_{\rm k}=E_{\rm f}/(cM_{\rm NS})\sim 10 E_{\rm f,45}\mbox{ cm s}^{-1}$. Over many flares, the contribution from the kicks would increase in a random walk process, eventually reaching $v_{\rm k,N}\approx 3\times 10^3 E_{\rm f,45} N_{5}^{1/2}\mbox{cm s}^{-1}$. This is still, however, much smaller than typical neutron star velocities and hence is not likely to be observable.

 \subsection{Spin-down due to mass-loaded charged particle winds}
 \label{sec:motivatePwind}
A sufficiently strong mass-loaded wind (with a kinetic luminosity $L_{\rm pw} \gtrsim L_{\rm dip}= B_{\rm dip}^2 R_{\rm NS}^6 \Omega^4/c^3$ where $L_{\rm dip}$ is the standard dipole spin-down luminosity, $B_{\rm dip}$ is the dipolar magnetic field, $R_{\rm NS}$ is the radius of the neutron star and $\Omega$ is its spin frequency) of charged particles can open-up field lines of a rotating magnetar beyond a radius of $R_{\rm open}\sim R_{\rm NS} (B_{\rm dip}^2 R_{\rm NS}^2c/L_{\rm pw})^{1/4}$ \citep{BT1998,Harding1999,Thompson2000}. This significantly enhances the spin-down, which scales as the amount of open magnetic flux squared, as compared with dipolar spin-down \citep{Bucciantini+06}. Under the influence of such a wind, $\dot{P}\propto P$ and the period increases exponentially
\begin{equation}
\label{eq:expP}
    P=P_0\exp (t/\tau) 
\end{equation}
on a growth timescale of
\begin{equation}
    \tau=\frac{I c R_{\rm open}^2}{B_{\rm dip}^2 R_{\rm NS}^6}=\frac{I c^{3/2}}{B_{\rm dip} R_{\rm NS}^3L_{\rm pw}^{1/2}}=5\times 10^7 B_{\rm dip,15}^{-1}L_{\rm pw,40}^{-1/2}\mbox{ sec}.
\end{equation}

A mass-loaded wind strong enough to open-up field lines in this manner is not expected to be operating at all times in the life of a magnetar.
However, using the radio afterglow of the GF from SGR 1806-20, \cite{Gelfand+05,Granot2006} have inferred the existence of a mildly relativistic outflow with a kinetic energy comparable to that of the GF. In addition, a mass-loaded wind with similar properties is needed to power the persistent synchrotron source which is spatially coincident with FRB 121102 assuming it is the birth nebula of a magnetar \citep{Margalit2018}.

If an order-unity fraction of a magnetar's internal energy goes into the kinetic energy of a mass-loaded wind, then the final spin period of the magnetar is given by
\begin{equation}
\label{eq:Pf}
P_{\rm f}\!=\! P_0 \exp\bigg[\frac{E_{\rm B} \Delta t_{\rm pw}}{E_{\rm f}\tau} \bigg] \! =\! P_0 \exp \bigg[0.7 \frac{B_{\rm int,16}^2 B_{\rm dip,15}E_{\rm pw,42}^{1/2}\Delta t_{\rm pw,2}^{1/2}}{E_{\rm f,44}} \bigg] \end{equation}
where $\Delta t_{\rm pw}$ is the duration of the particle-wind emission, $E_{\rm pw}\approx L_{\rm pw}\Delta t_{\rm pw}$ is its kinetic energy and $E_{\rm f}$ is the average energy of a GF.
 
Equation \ref{eq:Pf} shows that, given a fixed energy reservoir, it is more favourable for purposes of efficient spin-down to have a longer lived outflow, even if it is significantly less luminous as compared to the peak gamma-ray luminosities of GFs.  Supporting a longer-timescale outflow, GFs exhibit long-lasting, pulsating X-rays `tails' which release $\sim 10^{44}$ erg over a few hundreds of seconds \citep{Hurley2005}.
These tails exhibit a super-Eddington luminosity from a Compton-thick anisotropically-emitting `photosphere', and thus {\it necessitate} the existence of a mass-loaded wind (although the duration of the wind remains unclear) in order to advect the energy to large radii where it can be radiated \citep{VanPutten2016}. If the luminosity and timescale of the wind scale with the same properties for the radiation, the wind associated with tails dominates over a shorter lived outflow component associated with the GFs themselves. We therefore focus on this scenario as our canonical scaling below.
 
Equation \ref{eq:Pf} suggests that to attain ULPs via this process requires a magnetars born with a much stronger interior fields $B_{\rm int} \gtrsim 10^{16}$ G than the external dipole fields of most Galactic magnetars, $B_{\rm d} \sim 10^{14}-10^{15}$ G.  Although such strong birth fields are expected in some scenarios (such as from the core collapse of rapidly-rotating stars, which generate MHD-powered supernovae; e.g.~\citealt{Mosta+14}), and may be needed if millisecond magnetars are to power gamma ray bursts (\citealt{BGM2017} and references therein), they may not be generic.  On the other hand, the exponential sensitivity of $P_{\rm f}$ to the physical parameters suggests that if only a small fraction of magnetars are born with favourable conditions, they may attain spin periods orders of magnitude larger than would be obtained from ordinary dipole spin-down alone.
 
 \subsection{Proof of Concept: Monte Carlo Simulation}
\label{sec:MonteCarlopw}
As a proof of concept of the mechanism discussed above, we present here a Monte Carlo calculation of the period distribution of a population of magnetars arising from spin-down due to a mass-loaded wind (and thus of potential periodicity imprinted on their FRBs). Owing to uncertainties in the birth characteristics of magnetars, our calculation necessarily involves a few speculative assumptions.  Nonetheless, it reveals the general picture to be conceptually valid and provides some base level expectations.

Our calculation proceeds as follows.  We assume that magnetars are born with surface dipole magnetic field strengths, $B_{\rm dip,0}$, which are log-normally distributed with a median value of $10^{14.75}\mbox{ G}$ and a scatter of 0.5 dex.  The internal magnetic field, $B_{\rm int,0}$, is log-uniformally distributed between the dipole field value and a value ten times stronger.  We assume that the energy of the GFs varies according to $dN/d{\cal E}\propto {\cal E}^{-1.7}$ (consistent with an extension of the magnetar short burst distribution) with values ranging from ${\cal E}_{\rm min}$ to ${\cal E}_{\rm max}=E_{\rm int}$.
A minimum requirement on ${\cal E}_{\rm min}$ is that it provide a sufficiently luminous flare to overcome the magnetic Eddington limit, accounting for the suppression of the electron scattering cross section for photon energies well below the first Landau state.  This is a complex radiative transfer problem \citep[e.g.,][]{2013MNRAS.434.1398V,VanPutten2016} which depends on photon angles with respect to the local $B$ and differing polarization states. For expediency, we adopt an isotropic Rosseland mean approximation for the opacity deep in the fireball photosphere as considered by \cite{Paczynski1992}, in which case this constraint reads ${\cal E}_{\rm min}/t_{\rm f}\gtrsim 3.5 \times 10^{38}\max(1,B_{\rm dip,12})^{4/3}\mbox{ erg s}^{-1}$.

The time required for magnetic energy to leak from the magnetar interior to its surface in general depends on the strength of the magnetic field (and therefore on time). For example, if it is governed by ambipolar diffusion in the ordinary case of modified URCA cooling \citep{Beloborodov&Li16}, then one expects a magnetic power of $\dot{E}_B\approx 10^{39} B_{\rm int,16}^{3.2}$ erg s$^{-1}$ \citep{Margalit+19}. Once again taking $E_{\rm B}=B_{\rm int}^2 R_{\rm NS}^3/6$, this relation leads to a decay of the internal field as a function of time as
\begin{equation}
     B_{\rm int}(t)=B_{\rm int,0}[1+1.13t_{\rm yr,3}B_{\rm int,0,16}^{6/5}]^{-5/6}
\end{equation}
 where $t_{\rm yr,3}$ is the time since the formation of the magnetar in units of $10^3$yrs. The magnetic energy therefore decays on a typical timescale of $\tau_B= 880 B_{\rm int,0,16}^{-6/5}$ yr, which is consistent with the inferred histories of Galactic magnetars \citep{Beniamini2019}. 
 
 For simplicity we assume that at all times, the ratio of the internal and dipole field strengths remains the same.  After each GF, the period increases according to equation \ref{eq:expP} with $E_{\rm pw}=E_{\rm f}, \Delta t_{\rm pw}=300$ s. If the duration of the wind after a flare is longer than this, then the total flare energy required to achieve the same spin period would be reduced (see equation~\ref{eq:expP} and surrounding discussion). The time until the next GF, $T$, is then self-consistently accounted for, given the current magnetic energy loss and the energy of each GF according to $\int^T\dot{E}_{\rm B}dt=E_{\rm f}$ (this ensures that enough energy is supplied to the surface between flares to account for the next flare; it also implicitly assumes that GFs dominate the release of magnetic energy, which though uncertain is consistent with the Galactic magnetar population, see \citealt{Beniamini2019}). For $B_{\rm int} \approx 5\times 10^{14}$ G, this results in a time between flares of $10^{44}$ erg of $T\approx 25$ yr, broadly consistent with the interval between GFs of a given magnetar in the Galactic population.
 
 In between flares the magnetar is assumed to spin-down at the standard dipole rate, taking into account the decaying dipole field\footnote{ Our assumption that the ratio of dipole to internal field remains constant throughout the evolution is equivalent to a choice of $\alpha=1.2$ in the standard formulation $\dot{B}\propto B^{1+\alpha}$ \citep{Colpi2000}. Larger (smaller) values of $\alpha$ would lead to a slower (faster) decay of the magnetic field at $t>\tau_{\rm B}$ and therefore correspond with more (less) ULPs. }.
 Integrating the relation $d\Omega/dt'= -\Omega(t')^3 B_{\rm dip}(t')^2R^6/(c^3I)$ from $t$ to $t+T$, we find
 \begin{eqnarray}
& P(t+T)^2=P(t)^2+\frac{12\pi^2 R_{\rm NS}^6 B_{\rm dip}(t)^2\tau_B(t)}{Ic^3}(1-(1+x)^{-2/3})\\
& \mbox{ for } x\equiv T/\tau_B(t). \nonumber
 \end{eqnarray}
 We continue simulating flares until the magnetar has lost all of its initial magnetic energy (at which point it is no longer a magnetar). Other parameter values assumed in this calculation are, $P_0=0.01\mbox{ s}, R_{\rm NS}=10\mbox{ km}, M_{\rm NS}=1.4M_{\odot}, I=1.3 \times 10^{45} \mbox{\, g cm}^2$.
 
 \begin{figure}
\centering
\includegraphics[width=0.4\textwidth]{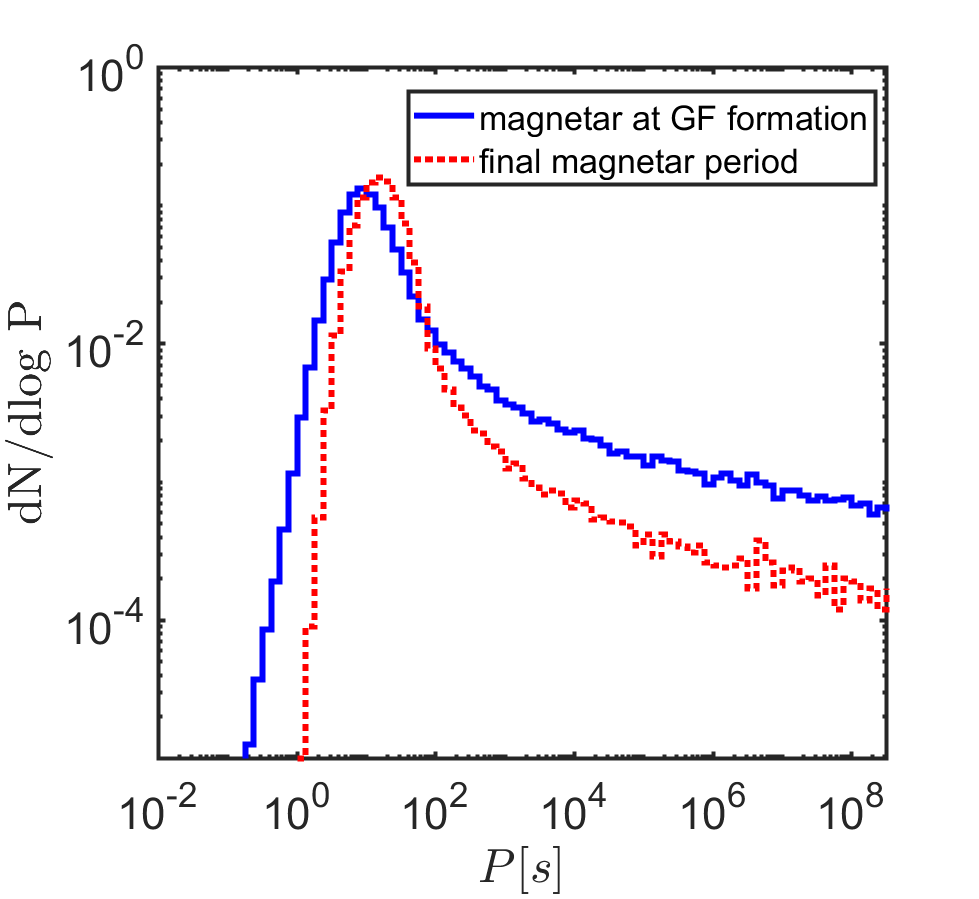}\\
\includegraphics[width=0.4\textwidth]{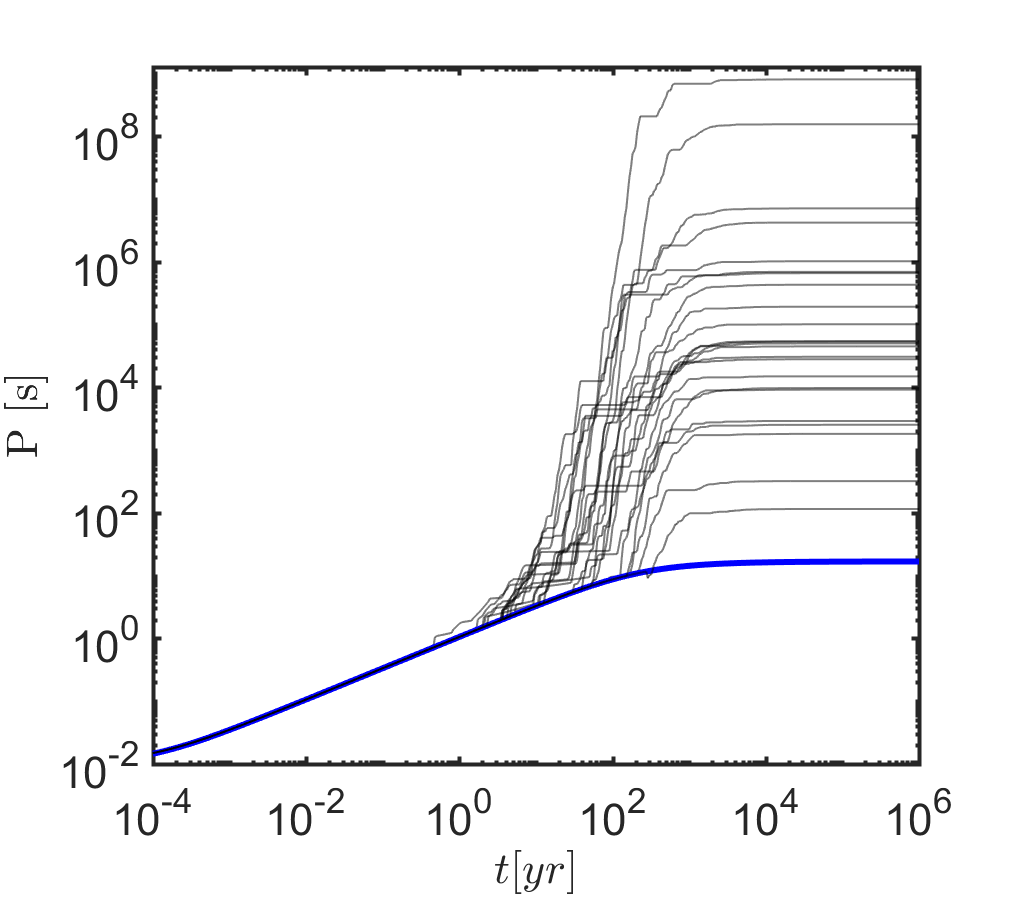}
\caption{Top: The final period distribution (dotted) of magnetars taking into account windows of enhanced spin-down due to the presence of a mass-loaded wind. The solid curve depicts the period distribution of magnetars at the time of GF production.
Bottom: Examples of the period evolution with time for a magnetar with $B_{\rm dip,0}=4\times 10^{15}\mbox{ G}$ and $B_{\rm int,0}/B_{\rm dip,0}=10$. A thick (blue) line denotes the evolution of a magnetar with the same initial conditions and no period evolution due to mass-loaded winds.
}\label{fig:MonteCarloPWspin}
\end{figure}

The top panel of figure \ref{fig:MonteCarloPWspin} shows the resulting final spin period distribution based on the Monte Carlo calculation described above with $3\times 10^4$ realizations as well as the distribution of periods associated with each of their GFs (which in some FRB models, e.g. \citealt{Metzger+19}, would correspond to the FRB period distribution). In the low-twist model, FRBs are associated with short bursts rather than GFs. However, since GFs are associated with intense short burst activity \citep{1999Natur.397...41H}, the periods associated with GFs may still be a good proxy for the magnetar periods associated with FRBs in that model.  Our results reaffirm the expectations from Galactic magnetars, that the period distribution should peak at $\sim 10$ s. However, we also find a tail of the population that extends to periods several orders of magnitude larger.  The fraction of magnetars with $P=2\times 10^4-10^6$ s (corresponding to 1E 161348-5055 and FRB 180916 respectively) is $\sim 6\times 10^{-3}$ of the entire population. 
The fraction of FRB-generating magnetars with such periods would exceed this fraction if FRBs are preferentially produced from slowly-rotating magnetars (see \S \ref{sec:LongPeriodlowtwist}).
In addition, this fraction is enhanced to $\sim 2\times 10^{-2}$ when considering each FRB to be associated with a single GF. This is because as the magnetars' age beyond $\sim \tau_B$, their period can no longer increase due to spin-down, and their period evolution is solely due to GF production. From this point onward, their flares tend to become less energetic, implying that (if their initial fields were sufficiently strong) they can produce an increasing number of GFs in a given logarithimic period range as time goes by (and their period increases).

The bottom panel of Figure~\ref{fig:MonteCarloPWspin} shows example evolutionary tracks for the spin period of magnetars with $B_{\rm dip,0}=4\times 10^{15}\mbox{ G}$ and $B_{\rm int,0}/B_{\rm dip,0}=10$, respectively.  If such strong birth magnetic fields are realized in nature, then periods as large as those seen in 1E 161348--5055 and FRB 180916 can be achieved within $\sim 10^3$ yrs of formation, consistent with the age of the supernova remnant RCW 103 hosting 1E 161348--5055.  For the same model parameters as given above, the magnetic field strength at the present epoch ($t\sim 3\times 10^3$ yrs) is $B_{\rm dip}\approx 3\times 10^{14}\mbox{ G}$, while the present-day period derivative is
\begin{equation}
\dot{P}=\frac{4\pi^2 B_{\rm dip}^2R^6}{Ic^3P}\approx 10^{-15} B_{\rm dip,14.5}^2 P_{5}^{-1}.
\end{equation}
The latter is consistent with the upper limit of $\dot{P} < 1.6 \times 10^{-9}$~s s$^{-1}$ for 1E 161348--5055 \citep{2011MNRAS.418..170E}.

\begin{figure}
\centering
\includegraphics[width=0.45\textwidth]{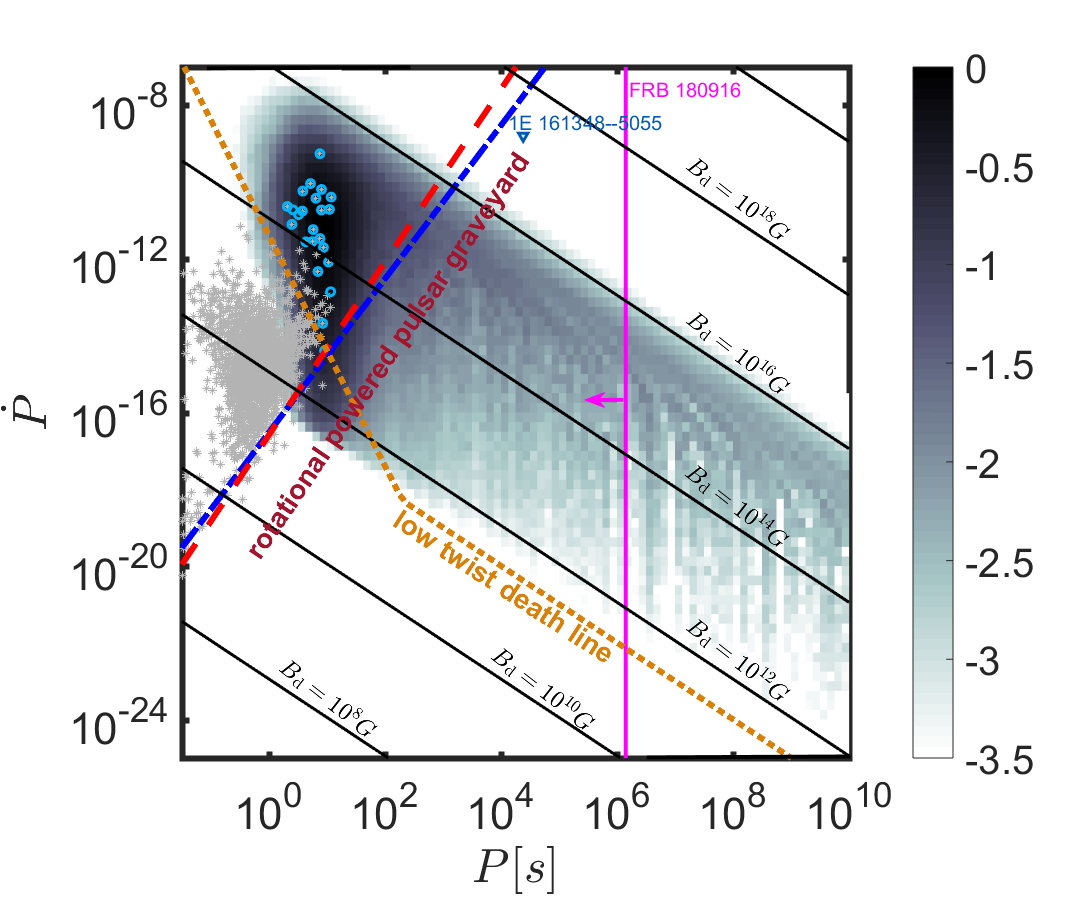}
\caption{Density of simulated magnetars (in $\log_{10}$ units, normalized to the peak) undergoing spin-down due to mass-loaded winds in the period - period derivative plane. Also shown are observed pulsars (gray stars), confirmed magnetars (light blue circles) as well as 1E 161348-5055 (blue arrow; with only an upper limit on $\dot{P}$) and FRB 180916 (pink solid line; no information on $\dot{P}$, and $P$ may be smaller due to period aliasing). Contour lines depict the dipole magnetic field lines. The approximate estimates for the pulsar death line are shown by dashed and dot-dashed lines (respectively cases I' and III' from \citep{Zhang2000}). Finally, the `death line' for FRB creation in the low twist model \citep{Wadiasingh2020} (hereafter \citetalias{Wadiasingh2020}) is depicted by a dotted line. Allowed sources lie above the line.}\label{fig:PPdotmap}
\end{figure}

Figure \ref{fig:PPdotmap} depicts the distribution of simulated magnetars in the $P-\dot{P}$ diagram, overlaid with observed pulsars (taken from the ATNF catalog\footnote{\url{http://www.atnf.csiro.au/research/pulsar/psrcat/}}  \citep{2005AJ....129.1993M} and confirmed magnetars. The ULP systems reside well beyond the conventional `pulsar death line' or `curvature radiation pulsar pair death line'\footnote{Recent simulations have demonstrated that pair cascades are nonstationary \citep{2010MNRAS.408.2092T,2013MNRAS.429...20T}, thus stationary gap calculations of death lines should not be over-interpreted. More precisely, the death line is actually a broad death band where pulsars become increasingly `weaker'.}, demonstrating that they would not be observable as classical radio pulsars.

\section{ULPs from fallback accretion}

The late-time accretion of supernova ejecta through a fallback disk\footnote{Fallback or fossil disks historically were a prominent alternative (and contentious) model for AXPs, and the topic was regarded with skepticism by the magnetar community.} \citep{1988Natur.333..644M} and its resulting torque on the central neutron star has long been considered a promising mechanism to explain the ULP of 1E 161348--5055 \citep{2007ApJ...666L..81L,2016ApJ...833..265T,2017MNRAS.464L..65H,2019ApJ...877..138X}, although most previous works on this subject make a number of overly simplistic assumptions. Spectroscopic calorimetry of RCW 103 with blast-wave models provide evidence for a sub-energetic supernova explosion \citep{2019MNRAS.489.4444B,2019A&A...629A..51Z}, consistent with a significant quantity of fallback material.

\cite{MBG2018} presented a model for fallback accretion onto millisecond magnetars, assuming that the magnetic field is aligned with the spin vector and taking into account the angular momentum and energy coupling between the disk and the magnetar, and focusing on the situation in which the fallback radius and the outer edge of the disk, are larger than the other characteristic radii in the problem. The full treatment of the fallback evolution at late times is rather complex and can involve many regimes (depending on the relative locations of all the critical radii in the problem as a function of time, e.g., the neutron star radius, $R_{\rm NS}$, the co-rotation radius, $R_{\rm c}$, the Alfven radius, $R_{\rm m}$, the light-cylinder, $R_{\rm lc}$, the fallback radius, $R_{\rm fb}$, and the outer edge of the spreading accretion disk, $R_{\rm out}$). We will return to this problem in greater detail in a future publication. In this work, we merely wish to demonstrate that it is possible in principle to obtain ULP magnetars due to fallback evolution.

Characterizing the fallback accretion rate at the inner accretion radius according to
\begin{equation}
    \dot{M}_{\rm in}=\dot{M}_{\rm i}\bigg(1+\frac{t}{t_{\rm fb}}\bigg)^{-\zeta}
\end{equation}
and considering $t\gg t_{\rm fb}$, a steady state spin-down is typically obtained, in which the accretion spin-up is balanced by torques from the star-magnetosphere interaction such that the Alfv\'{e}n radius is slightly above the co-rotation radius ($R_{\rm m}\approx R_{\rm c}$).
Typically, the inner edge of the accretion radius is governed by the Alfv\'{e}n radius ($R_{\rm m}\propto \dot{M}_{\rm in}^{-2/7}$ see \citealt{1975A&A....39..185I,Ghosh1978}), below which the disk is disrupted. So long as this equilibrium is maintained, the magnetar's spin-down is modified as compared to the case of standard dipole spin-down, according to $\Omega\propto t^{-3\zeta/7}$. This implies an enhanced spin-down, as compared to the standard dipole case for an isolated magnetar (with $\Omega\propto t^{-1/2}$) for $\zeta>7/6$.

One can show that a minimum value of $\zeta$ is required to spin-down the magnetar to $P_{\rm obs} \gtrsim 10^6$ s.  As a best-case limiting scenario, consider that a steady-state evolution can be obtained on a timescale shorter than the initial fallback time, $t_{\rm fb}$, over which time the accretion rate is approximately constant (i.e. $t_{\rm c}<t_{\rm fb}$ for the definition of $t_{\rm c}$ shown below).  The spin period at this equilibrium state is given by equating $R_{\rm m}$ and $R_{\rm c}$ \citep{MBG2018}, leading to
\begin{equation}
    P_{\rm c}=11B_{\rm dip,16}^{6/7}\dot{M}_{\rm i,-2}^{-3/7}M_{1.4}^{-5/7}\mbox{ ms}
\end{equation}
 where $M_{1.4}$ is the NS mass in units of $1.4M_{\odot}$. When the initial period is smaller than $P_{\rm c}$, and while $R_{\rm NS}<R_{\rm c}<R_{\rm m}<R_{\rm lc}$ the evolution towards $P_{\rm c}$ increases exponentially, on a timescale
 \begin{eqnarray}
  &   t_{\rm c}=\frac{I}{\frac{B^2R^6}{cR_{\rm m}^2}+\dot{M}_{\rm i}R_{\rm m}^2}\approx \\& \bigg(0.11B_{\rm dip,16}^{6/7}\dot{M}_{\rm i,-2}^{4/7}M_{1.4}^{-17/14}+1.05B_{\rm dip,16}^{8/7}\dot{M}_{\rm i,-2}^{3/7}M_{1.4}^{-25/14}\bigg)^{-1}\mbox{ s} \nonumber
 \end{eqnarray}
 where the first term in the denominator is the torque due to enhanced spin-down when $R_{\rm NS}<R_{\rm m}<R_{\rm lc}$ (see \citealt{2016ApJ...822...33P,MBG2018}) and the second term is the toque due to accretion, at the limit $R_{\rm m}\gg R_{\rm c}$ \citep{Piro2011}.
 The time it takes the magnetar to reach a given period $P_{\rm c}$ much greater than its initial one is, up to a logarithmic factor $\Lambda \equiv \log(P_{\rm c}/P_0)$, equal to $t_{\rm c}$. The period remains at $P_{\rm c}\sim$ const so long as $\dot{M}\sim$ const. Once $\dot{M}$ starts decreasing (at $t=t_{\rm fb}$), $\Omega$ also decreases according to the equilibrium $R_{\rm m}\sim R_{\rm c}$. Assuming the steady state evolution is maintained until late times when the magnetic field decays (this requires a sufficiently large outer extend of the disk, a point we shall return to below), the magnetar's final spin period (i.e. at the time, $\tau_{\rm B}$, of magnetic field decay, see \S \ref{sec:MonteCarlopw} for details) is given by
\begin{equation}
    P_{\rm f}=P_{\rm c}\bigg(\frac{\tau_{\rm B}}{t_{\rm fb}}\bigg)^{3\zeta/7}
\end{equation}
Requiring $P_{\rm f}=P_{\rm obs}$, implies a lower limit on the value of $\zeta$
 \begin{eqnarray}
 \label{eq:limzeta}
 &    \zeta>\zeta_{\rm cr}\approx \frac{7}{3}\frac{\log\left[9\times 10^7M_{1.4}^{5/7}\dot{M}_{\rm i,-2}^{3/7}P_{\rm obs,6}B_{\rm dip,16}^{-6/7}\right]}{\log\left[2.7\times 10^9 B_{\rm dip,16}^{-6/5}t_{\rm fb,1}^{-1}\right]}\approx 1.96
 \end{eqnarray}
 where we have assumed here an internal magnetic field strength equal to the external dipole field.  Evidently, the standard fallback rate with $\zeta=5/3$ \citep{1988Natur.333..644M} decays too shallowly to achieve the large observed periods.  
 
However, a large value of $\zeta$ is possible in some cases.  For instance, at such high accretion rates the disk may be unable to cool efficiently, instead forming a radiatively-inefficient accretion flow characterized by powerful winds that carry away mass from the disk as matter flows to smaller radii.  In particular, the accretion rate reaching some inner radius $r_{\rm in}$ depends on the rate at the outer feeding radius ($r_{\rm out}$) as
 \begin{equation}
 \label{eq:Mdotp}
    \dot{M}_{\rm in}=\dot{M}_{\rm out} \bigg(\frac{r_{\rm in}}{r_{\rm out}}\bigg)^p, 
 \end{equation}
where $0\leq p \leq 1$ \citep{Blandford1999}.
We note however that this expression may not hold throughout the disk if, for example, the outer edges of the disk are no longer super-Eddington at late times, which is not a trivial specification \citep{Margalit2018}. The requirement of locally super-Eddington conditions throughout the disk at all times may be relaxed somewhat if, for instance, the inner parts of the disk are able to launch an outflow that would ablate matter and reduce the accretion rate from the outer disk, similar to feedback from Active Galactic Nuclei (e.g. \citealt{Fabian2012}).
If the outer edge of the disk is that of a viscously-spreading torus, conservation of angular momentum results in $r_{\rm out}\propto t^{2/3}$ \citep{Metzger2008}. Assuming the accretion rate at the outer edge of the disk is dominated by the spreading of the initial disk mass, $\dot{M}_{\rm out}\propto t^{-(2p+4)/3}$ \citep{Metzger2008}. Plugging this back into equation \ref{eq:Mdotp} we find
\begin{equation}
\label{eq:zetawithp}
  \dot{M}_{\rm in}=\dot{M}(R_{\rm m})\propto t^{-\zeta} \quad ; \quad   \zeta=\frac{28(p+1)}{3(7+2p)}
\end{equation}
The value of $\zeta$ is maximized for $p=1$ at a value of $2.07$, somewhat larger than the required value $\zeta_{\rm crit}$. Although a large number of uncertainties have entered this calculation, this suggests that it is potentially possible that fallback accretion could significantly increase the magnetar period, up to those measured for 1E 161348--5055 ($P_{\rm obs}=10^{4.4}$ s) and even FRB 180916 ($P_{\rm obs}=10^{6.15}$ s).

The conditions outlined above impose constraints on the initial fallback conditions. One such constraint is that the fallback should persist until late times, on the order of the magnetar's active lifetime $\tau_{\rm B}$.
At minimum, this requires that the outer edge of the disk, $R_{\rm out}$, always lies above the inner edge at $R_{\rm in}=R_{\rm m}$.
At times much greater than the fallback time, $t_{\rm fb}$, and the viscous time at the outer edge of the initial disk, $t_{\rm visc}(R_{\rm fb})$, the outer radius of the disk evolves as $R_{\rm out}\propto t^{2/3}$. At the same time, the asymptotic expansion of the inner edge is given by $R_{\rm m}\propto t^{2\zeta/7}$. Since $\zeta\lesssim 2.1$, the asymptotic growth of $R_{\rm out}$ is always faster than that of $R_{\rm in}$. The limiting condition for the existence of the disk at late times, is therefore given by the initial setup, i.e. the requirement is that the initial fallback radius satisfies $R_{\rm fb}>R_{\rm m,0}$,
\begin{equation}
    R_{\rm fb}>8\times 10^6 B_{\rm dip,16}^{4/7}\dot{M}_{\rm i,-2}^{-2/7}M_{1.4}^{-1/7}\mbox{ cm}
\end{equation}
or equivalently, in terms of the angular momentum per unit mass, $j_{\rm fb}\gtrsim 4\times 10^{16}\mbox{ cm}^2\mbox{ s}^{-1}$.

\begin{figure}
\centering
\includegraphics[width=0.4\textwidth]{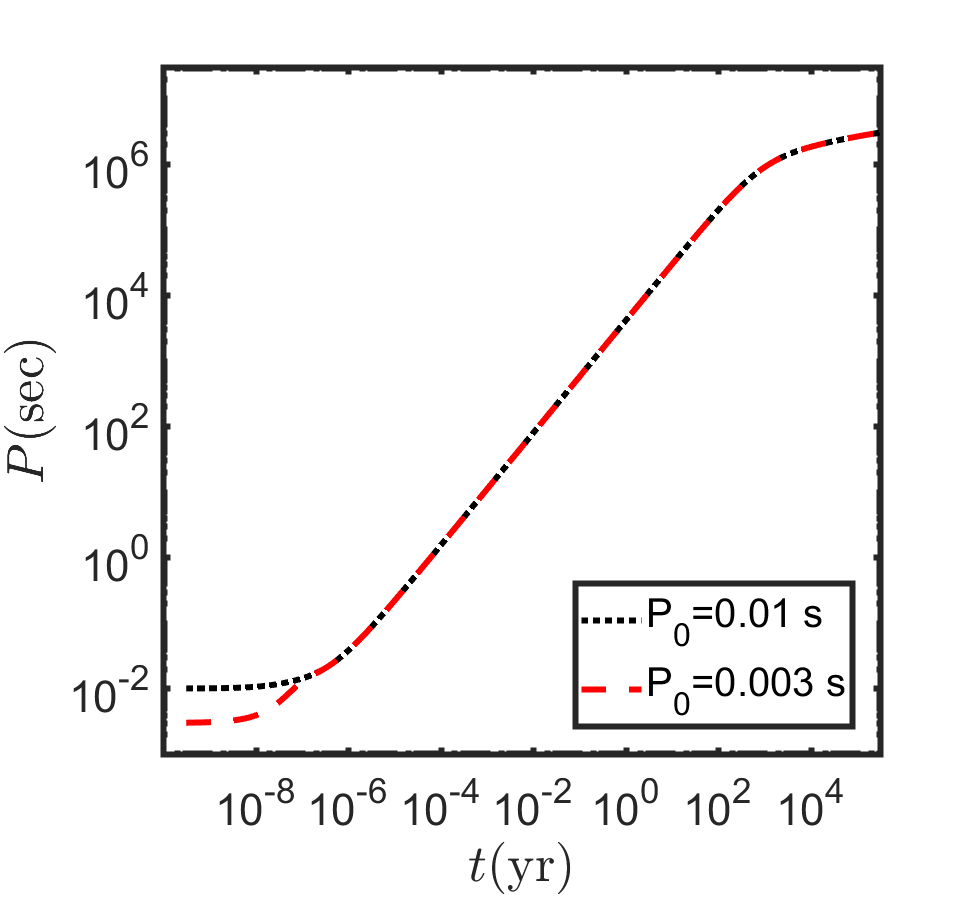}\\
\includegraphics[width=0.4\textwidth]{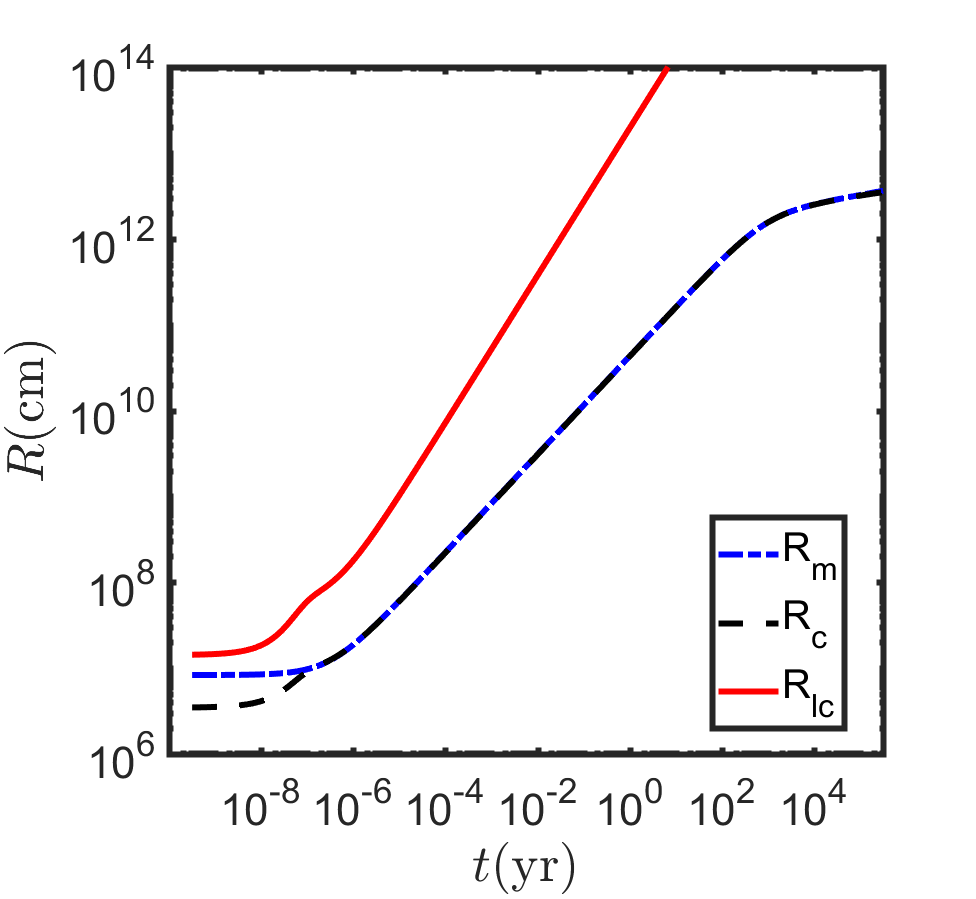}
\caption{Top: Evolution of a magnetar's period due to fallback accretion, with $B_{\rm int}=B_{\rm dip}=10^{16}\mbox{ G}, M_{\rm fb}=0.1M_{\odot}, t_{\rm fb}=10\mbox{ s}, \zeta=2$ and different values of the initial spin period $P_0$. Bottom: Evolution of the characteristic radii (Alfven radius, $R_{\rm m}$, co-rotation radius, $R_{\rm c}$ and the light-cylinder radius, $R_{\rm lc}$) using the same initial conditions (and the lower value of the spin period).}\label{fig:fallback}
\end{figure}

An additional condition that is required for equation \ref{eq:zetawithp} to hold is that the mass accretion rate is dictated by the initial mass of the spreading disk, and that the material falling onto the disk at times $t\gg t_{\rm fb}$ can effectively be ignored. This also appears reasonable, assuming that soon after $t_{\rm fb}$, $R_{\rm m}$ expands beyond $R_{\rm fb}$. In that case, matter attempting to falling back to the disk at $R_{\rm fb}$ at later times, has a lower specific angular momentum than matter at $R_{\rm m}$. A significant fraction of this matter could then be thrown off from the system by a propeller mechanism. It will likely acquire angular momentum in the process and in doing so, help spin-down the neutron star even faster. Ignoring this late fallback is therefore expected to be conservative for purposes of arguing for longer spin periods.

An example of a magnetar's period evolution due to fallback that can reach large periods of $\sim 10^6$ s is depicted in figure \ref{fig:fallback}. As evident from the figure, the final period does not depend strongly on the initial value (this is true so long as the period $P_{\rm c}$ can be obtained at a time shorter than $t_{\rm fb}$ as explained above). The magnetar's period increases as a power-law over many orders of magnitude in time, until $\sim 10^3$ yrs, when the magnetic field decays. As this point, the magnetic field decays sufficiently that the magnetar can no longer spin-down and the period effectively freezes.

\section{ULPs Favor FRB Production in Low-Twist Magnetar Model}
\label{sec:LongPeriodlowtwist}
We turn next to discuss some implications of the ULP magnetar period population on a specific model for FRB production. In particular, we demonstrate that ULP magnetars can provide fertile grounds for FRB production, and that such systems may overproduce FRBs as compared to magnetars with canonical spins.

In the low-twist magnetar model proposed in \citet{2019ApJ...879....4W}, long-lived states of low charge density in magnetars are unstable to avalanche magnetic pair production by small perturbations of the local magnetic field. These field dislocations are assumed to be triggered by the same underlying mechanism as recurrent magnetar short bursts. The minimum charge density required for FRB viability is gated by any persistent low-twist or the corotational Goldreich-Julian charge density, $\rho_{\rm burst} \gtrsim \max(\rho_{\rm corot}, \rho_{\rm twist})$. Here, $\rho_{\rm burst} \sim \xi \nu B/(2 \lambda c)$, where $B$ is the local magnetic field, $\xi$ the dislocation amplitude, and $\lambda, \nu$ the characteristic wavelength and frequency of perturbations (ascribed to crustal oscillations). As the twist could be low, the minimum charge density is set by corotation.

Without regard to a persistent charge density, the minimum amplitude $\xi$ is that which is required {{\it kinematically}} for avalanche magnetic pair production in a magnetar-like field. From equation ~(A11) of \citetalias{Wadiasingh2020}, the amplitude must be greater than
\begin{eqnarray}
\xi &\gg&  \frac{98}{9 \pi} \left(\frac{7}{3}\right)^{1/3} \lambda \left( \frac{\lambar^2 \rho_c^2 c^3 }{\delta R_*^7 \nu^3} \right)^{1/3}  \left(\frac{B_{\rm cr}}{B} \right) \label{eq:A11}\\
&\sim&  10^{-2} \, \, \rho_{c,7}^{2/3} \lambda_{5.5} \delta R_{*,6}^{-7/3} \nu_2^{-1} \left(\frac{B_{\rm cr}}{B} \right)   \quad \rm cm. \nonumber
\end{eqnarray}
for efficient low-altitude pair cascades. Here, $B_{\rm cr} = m_e^2 c^3/(\hbar e) \approx 4.4 \times 10^{13}$ G is the quantum critical field, $\lambar$ is the reduced electron Compton wavelength, $\delta R_* \ll 10^6$ cm is the maximum gap size, and $\rho_c \sim 10^7$ cm is the local curvature radius of field lines.

\begin{figure}
    \centering
    \includegraphics[width=0.45\textwidth]{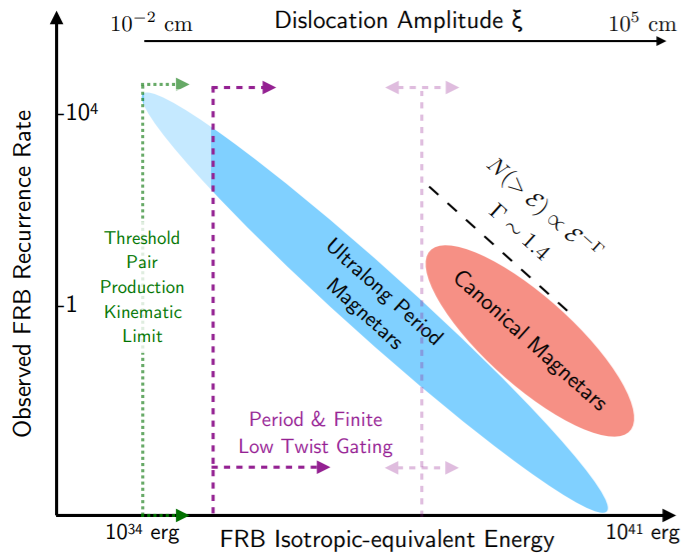}
    \caption{Schematic diagram of the FRB energy distribution in the low-twist magnetar model with FRB energy scaling as voltage caused by dislocations $\xi$. Shown in blue and red are two possible magnetar progenitor channels for FRB production, the ULP and canonical magnetars, respectively. At low energies, the ULP magnetars may dominate the global FRB energy distribution. The overall low-energy cutoff of individual repeaters is gated by their individual spin $P$ and state of low-twist and the global energy distribution reflects an unknown weighting of both populations/channels. }
    \label{schematic_fig}
\end{figure}

Interestingly, this kinematic characteristic minimum scale for $\xi$ is also consistent with charge starvation and energy conservation requirements for a ULP magnetar. For $P \sim 10^6$ s, the corotation charge starvation constraint is,
\begin{equation}
\xi \gtrsim \xi_{\rm min, ULP} = \frac{2 \lambda}{\nu P} \approx 6 \times 10^{-3} \frac{\lambda_{5.5}}{\nu_{2} P_{6}}\quad \rm cm
\label{ximinP}
\end{equation}
which is orders of magnitude smaller than for canonical magnetars where $\xi~\gtrsim~\xi_{\rm min, can} \sim~10^{2}$~cm. 

Since the number of perturbations of small amplitude $\xi$ greatly dominate those at larger $\xi$, the parameter space available for FRB production is enhanced with longer $P$, provided that the magnetar attains local low-twist $ \Delta \phi \lesssim 10^{-8} P^{-1}_{6}$. The differential number of short bursts at a given energy/fluence $dN/d{\cal E}_{\rm sb} \propto {\cal E}_{\rm sb}^{-s}$ has a well known index $s\approx 1.7$ in Galactic magnetars. Under the assumption ${\cal E}_{\rm sb} \propto \xi^2$, the lowest-energy bursts  allowed kinematically release energy $ 10^{-10}E_{{\rm max},44}~\sim~10^{34}$ erg. In the low-twist magnetar model for FRBs, we relate the familiar galactic magnetar short burst index to the distribution of voltage drops for pair cascades, and consequently FRB energies\footnote{This is guided by the observed FRB fluence distribution in repeaters FRB 121102 and FRB 180916, both of which exhibit $N \propto {\cal E}_r^{-1.5} $ and the detail that the radio luminosity in pulsars seems to scale as the polar cap voltage drop \citep{2002ApJ...568..289A}.} $dN/d{\cal E}_r \propto dN/d\xi \propto \xi^{-(2s-1)}$ such that $N(>{\cal E}_r) \propto \xi^{-2(s -1)}\approx \xi^{-1.4}$.

For a beaming fraction $ f_{b} \sim 0.1$ (suggested by the phase width of FRB 180916) and other characteristic scales promulgated in \citetalias{Wadiasingh2020}, energy conservation (equation~(15) in \citetalias{Wadiasingh2020}) requires
\begin{eqnarray}
&& \xi \gtrsim  \xi_{\rm min, ULP} \gg \xi_{\rm max} \left( \frac{{f_b} {\cal E}_0}{ E_{\rm max}} \right)    \left( \frac{B}{B_r} \right)  \label{efficiencyconstraint} \\
&&\sim  10^{-5} \xi_{\rm max}  f_{b,-1} {\cal E}_{0, 40}   E_{{\rm max},44}^{-1} \left( \frac{B}{B_r} \right) \sim 10^{-0.5} \xi_{\rm max, 4.5} \rm \quad cm \nonumber.
\end{eqnarray}
Here, a linear scaling of FRB energy with $\xi$ is assumed with the $B_r \sim 10^{15}$ G and  ${\cal E}_0 \sim 10^{40}$ erg normalization constants in the \citetalias{Wadiasingh2020} energy distribution obtained via the inversion protocol expounded in that work. Thus, we regard $\xi_{\rm min}~\sim~10^{-2} - 10^{-1}$~cm as the low-amplitude limit for FRB tenability.  

 Consequently, for other factors constant such as the rarity of low-twist states, the burst rate for a ULP magnetar over a canonical magnetar is increased by a factor $\sim (\xi_{\rm min, can}/\xi_{\rm min, ULP})^{2(s -1)}~\sim~10^4$, at lower luminosities. This implies such ULP magnetars dominate the low-end of the FRB energy distribution unless canonical magnetars outnumber ULP magnetars by a factor greater than  $\sim 10^4$ -- see figure~\ref{schematic_fig}.

For a proof-of-concept study of the broadening of the energy distribution from individual repeaters and consequently a population of uncommon ULP magnetars, we adopt a similar protocol to that described in \S4 of \citetalias{Wadiasingh2020} which assumes all FRBs are repeaters and arise from low-twist magnetars. For caveats and limitations of such an exercise, see \citetalias{Wadiasingh2020}. Instead of population distributions corresponding to canonical magnetars, we adopt the resulting distribution from the wind-braking simulations (figures~\ref{fig:MonteCarloPWspin}-\ref{fig:PPdotmap}). As in \citetalias{Wadiasingh2020}, for a fixed $P$ and characteristic surface $B$, for each amplitude $\xi$ drawn from a distribution $dN/d\xi \propto \xi^{-2.4}$,  we assign an isotropic-equivalent energy (regarded for each FRB pulse) linearly scaled to the amplitude assuming that the conditions associated with kinematic viability equation~(\ref{eq:A11}), charge starvation equation~(\ref{ximinP}), energy conservation equation~(\ref{efficiencyconstraint}), and magnetic confinement (equation~(7) in \citetalias{Wadiasingh2020}) are all met. We adopt the values $\xi_{\rm max} = 10^{4.5}$ cm, $\lambda = 10^{5.5}$ cm, ${\cal E}_0 = \{10^{41}, 10^{42}\}$ erg,  $B_r = 10^{15}$ G, $E_{\rm max} = 10^{44}$ erg, $f_{b} = 0.1$, $\nu = 100$ Hz, $\delta R_* = \{10^5, 10^6\}$ cm,  and $\rho_c = 10^7$ cm (we also set $\Delta R_* = 10^5$ cm, see \citetalias{Wadiasingh2020} for details). A key difference from \citetalias{Wadiasingh2020} (other than $f_b \neq 1$) is that much smaller amplitudes are accessible and so the kinematic condition equation~(\ref{eq:A11}) can be important, particularly at higher values of $B$ or $\delta R_*$. In contrast, the conservation constraint given by equation~(\ref{efficiencyconstraint}) is operational at lower values of $B$. As a result, the lowest viable amplitudes are $\sim 10^{-7} \xi_{\rm max}$. This requires much larger samples to adequately probe the full inertial range of amplitudes than the significantly narrower simulated energy distributions in \citetalias{Wadiasingh2020}.

\begin{figure}
\centering
\includegraphics[width=0.45\textwidth]{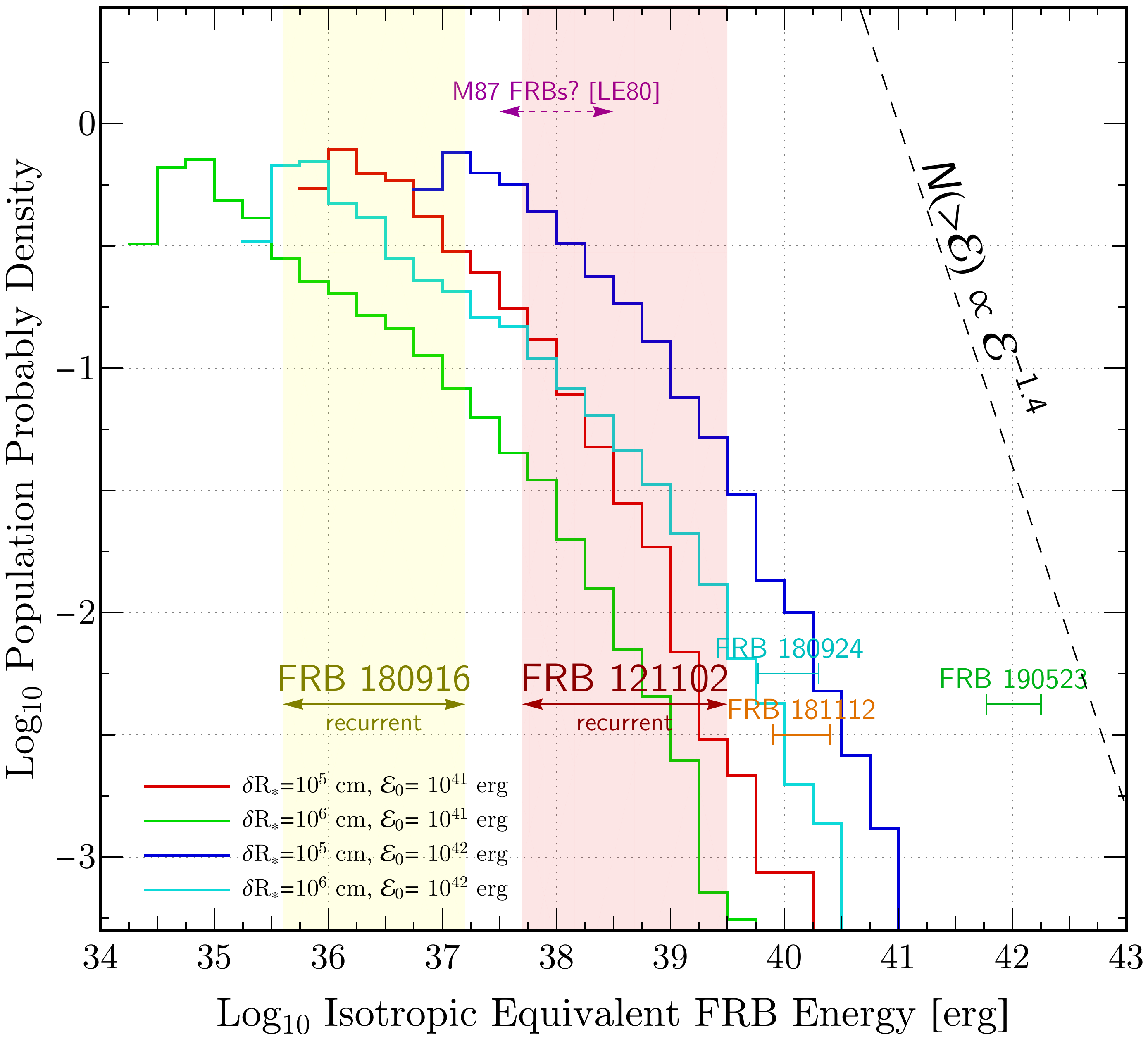}\\
\includegraphics[width=0.45\textwidth]{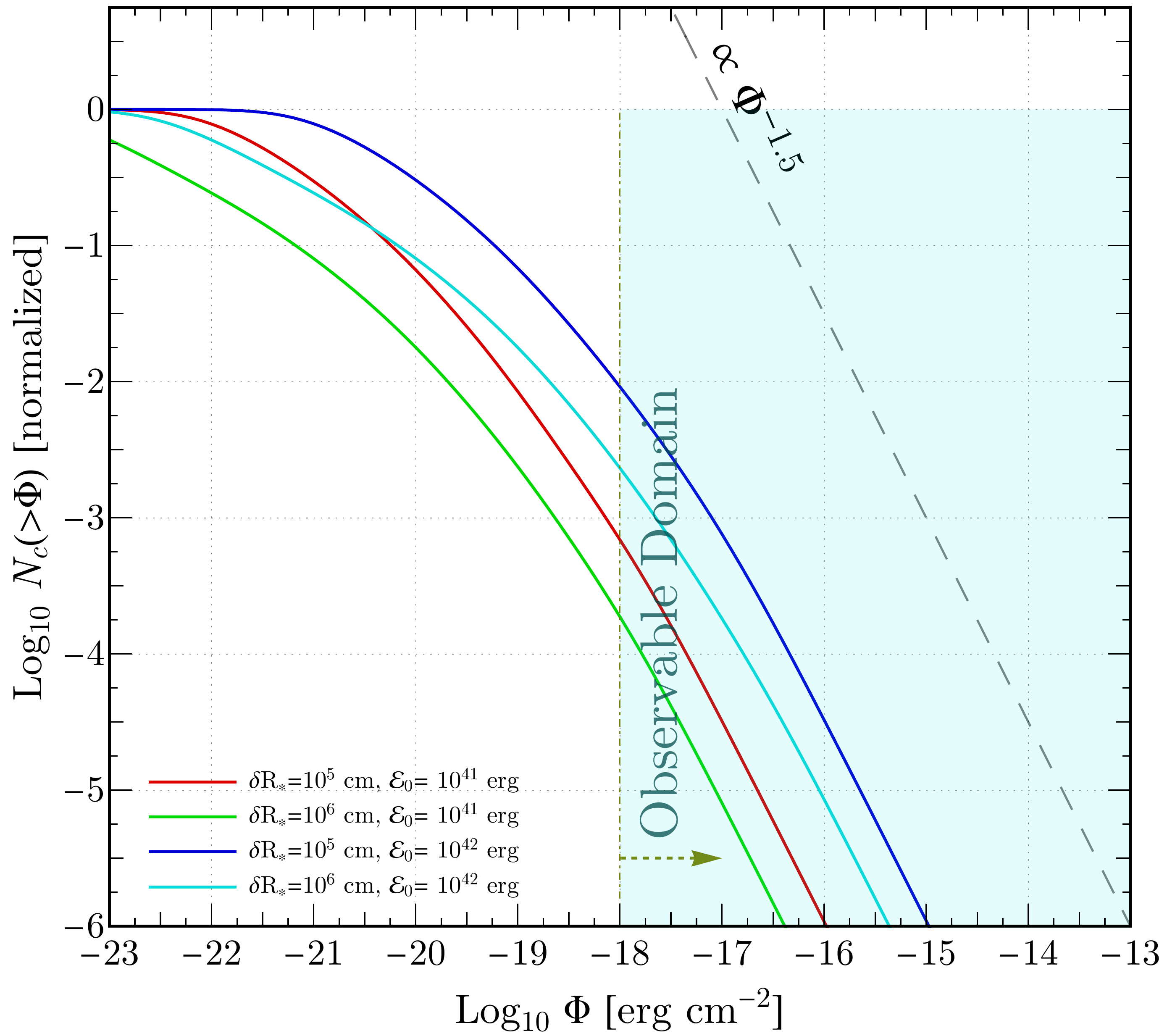}
\caption{Top: Ensemble population energy distribution realizations for the low-twist magnetar model adopting the population distribution from \S \ref{sec:MonteCarlopw}.
Bottom: Local fluence distributions for the corresponding curves of the top panel, assuming the magnetars follow the star formation rate in standard $\Lambda$CDM cosmology. The observable domain corresponds to a characteristic sensitivity of $\sim 100$ Jy $\mu$s for a flat spectral index in a bandwidth of $1$ GHz. 
}\label{fig:model_FRB_lumfuncs}
\end{figure}

Having constructed a list of energies for each individual repeater, we sample the population $dN/(d\log P d\log B )$  corresponding to  figures~\ref{fig:MonteCarloPWspin}-\ref{fig:PPdotmap} for $P$ and $B$ to construct an ensemble FRB energy distribution for a large population of repeaters. Displayed in the top of figure~\ref{fig:model_FRB_lumfuncs} are some realizations of such population distributions for four illustrative values of $\delta R_*$ and ${\cal E}_0$. All the curves in figure~\ref{fig:model_FRB_lumfuncs} are significantly flatter (and wider) than those in \citetalias{Wadiasingh2020} owing to the contribution of rare ULP magnetars. Looser restrictions on $\delta R_*$ (corresponding to larger values) broaden and flatten the population energy distribution as more lower amplitude disturbances can result in FRBs. The $\alpha = 1$ in \citetalias{Wadiasingh2020} is the most appropriate point of comparison, since the field evolution assumed in constructing figure~\ref{fig:PPdotmap} is $\alpha \approx 1.2$. Empirically, the wider character of the energy distributions are more consistent with observed isotropic-equivalent energies of localized repeaters FRBs and the brighter apparent non-repeaters FRB 190523, FRB 180924, and FRB 181112 \citep{1980ApJ...236L.109L,Tendulkar+17,Bannistereaaw5903,2019Natur.572..352R,Prochaska231,Marcote2020}. As in \citetalias{Wadiasingh2020}, at the very highest energies the distribution follows $N(>{\cal E}) \propto {\cal E}^{-1.4}$ since at these energies correspond the largest bursts in the highest-$B$ magnetars where period gating is not consequential.

In the bottom panel, we compute the respective local $N(>\Phi)$ distribution (``Log N -- Log S") distributions for local fluence $\Phi$  adopting standard $\Lambda$CDM cosmology and assume magnetars follow the star formation rate of \citep{2014ARA&A..52..415M} peaking at $z\approx 2$. Owing to the broad energy distributions in the top panel, the departures from standard Euclidean delta-function expectations ($N(>\Phi) \propto \Phi^{-1.5}$) are more pronounced than the narrower distributions in \citetalias{Wadiasingh2020}. However, these departures are at lower fluences than typically accessible by current facilities owing to the energy distributions in top panel peaking at lower energies. Since the low-energy cutoff in the distribution is $10^{35}-10^{36}$ erg in the top panel, the corresponding FRB rate in bottom panel implies the observed rate may increase substantially with future improvements in sensitivity.

 \section{Discussion}
The vast majority of observed Galactic magnetars have spin periods of $P\sim 1-10$ s. Their spin frequency decreases over time due to electromagnetic spin-down. Since their magnetic field decays on a timescale of $\sim 10^4$ yr, their spin-evolution eventually freezes, reaching a maximum period of $\sim 10$ s (see, e.g. \citealt{2013MNRAS.434..123V,Beniamini2019}). However, this evolution can be altered significantly under certain conditions. Motivated by the reported $16$-day periodicity in FRB 180916 and the 6.67 hr magnetar 1E 161348--5055 in RCW 103, we have explored plausible physical channels to attain long $P\gg 10$ s spin periods in magnetars. Our model is an alternative to previous magnetar models which instead invoke spin-precession or binary shrouding to account for FRB 180916's periodicity.

We scrutinize three mechanisms which may transfer angular momentum away from a magnetar as it matures: mass-loaded winds or kicks from bursts and interactions from supernova fallback. In either channel, a large $P \sim 10^4 - 10^6$ s implies: (a) an initially high internal magnetic field, $B_{\rm int}\gtrsim 10^{16}\mbox{ G}$, (b) a relatively mature magnetar with an age of $1-10$ kyr and (c) that ULP magnetars are rare compared to canonical magnetars of $P \sim 1- 10$ s, as required by observations of the Galactic systems. 
 
In the mass-loaded wind channel, bursts decrease the effective light cylinder radius and open-up field lines, enhancing the spin-down by causing episodic phases of exponential increase in period. We find that if GFs are typically followed by hundreds of seconds long mass-loaded winds, a continuum of periods is realized with a peak at $1-10$ s and a long tail of ULP magnetars comprising up to $\sim 0.1-1$\% of the total active magnetar population. GFs may also be followed by angular momentum kicks, but we find that those are likely to be sub-dominant compared to the mass-loaded winds. 

For the fallback scenario, a statistical estimate is hindered due to the inherent uncertainties in the supernova energetics and geometry which lead to uncertainty in the resulting fallback properties and, e.g., the potential development of a radiatively-inefficient accretion disk. That is, the distribution of initial conditions in the birthing supernova are poorly understood -- for instance, bimodality is suggested in young pulsar population kick velocities \citep{2002ApJ...568..289A}. In some ways, bimodality in the magnetar period distribution is more natural in the fallback scenario since it can be transmitted via a corresponding bimodality in either the supernova properties (i.e. amount of fallback or kicks) or existence of a binary companion prior to supernova.

Finally, although not considered in this work, fallback, episodic mass-loaded winds and angular momentum kicks may operate in parallel or even synergistically over a magnetar's lifetime, particularly if the magnetic obliquity is low. This would relax the required conditions for each channel on its own.
 
If most FRBs originate from repeating sources, with a burst energy distribution similar to that of FRB 121102, then the product of the volumetric birth rate of FRB-producing sources, $\mathcal{R}$, and their active timescale $\tau$ is constrained to be $\mathcal{R}\tau \sim 130(\eta \Phi)^{-1}$ Gpc$^{-3}$, where $\eta$ is the beaming factor and $\Phi$ the active duty cycle \citep{Nicholl+17}.  The total rate of core collapse supernovae in the local Universe is $2.5\times 10^{5}$ Gpc$^{-3}$ yr$^{-1}$, while at least $\sim 20\%$ of neutron stars are born as magnetars \citep{Beniamini2019}.  If FRBs are active for a timescale comparable to the ages of active Galactic magnetars $\tau \sim 10^{3.5}$ yr, then we conclude that at most a fraction $\sim 10^{-4}(\eta/0.1)^{-1}(\Phi/0.1)^{-1}$ of magnetars contribute to the repeating population. This supports the notion that only a small minority of magnetars need to evolve to ULPs as suggested in this work. This is also consistent with the recent analysis of \cite{Margalit2020} based on the observations of the first Galactic FRB, showing that the population of Galactic magnetars cannot simultaneously account for the observed rate and activity level observed in cosmological FRBs, and a second population, consisting of a small fraction of highly active magnetars can resolve these apparent inconsistencies.

Moreover, in some models of FRBs from magnetars (namely the low-twist model, \citealt{2019ApJ...879....4W}; \citetalias{Wadiasingh2020}), ULPs are highly favored for FRB production owing to the instability of the magnetosphere (with low persistent corotational charge density) to magnetic $e^+e^-$ pair cascades by disturbances caused by magnetar-like activity near the surface. In that model, a ULP magnetar fraction of $0.1-0.01$\% may be sufficient to dominate FRB production over that of canonical magnetars, particularly at lower isotropic-equivalent FRB energies. That is, most repeating FRBs may be from ULP magnetars while some apparently rarer non-repeating bursts at higher energies may arise from canonical magnetars.

\subsection{Predictions}

For the foreseeable future, observations of recurrent FRBs are likely to be restricted to the radio band.  Nevertheless, different models for burst periodicity (e.g. precession, binary shrouding and ULP scenarios) may be distinguishable given additional observations of FRB 180916 and the greater repeating FRB population.

\subsubsection{Individual Repeaters}

For FRB 180916 and future periodic FRBs, our model requires that beaming of FRB emission is associated with the poloidal magnetic axis. If the radio emission originates from the magnetosphere, frequency selection of bursts could be operating similar to pulsar radius-to-frequency mapping. 

We predict only a single periodicity, associated with the spin. In contrast, in both the precession and binary shrouding models, shorter periods of $1-10$ s  due to the magnetar spin may exist if the emission is beamed. Since the emission is necessarily beamed in precession models, shorter periodicity could be falsified with future prolific burst storms of FRB 180916 (and may be in tension with the non-detection of burst periodicity from FRB 121102). 

Since in the ULP scenario observed periodicity is associated with spin, residual dispersion measure ($\Delta$DM; beyond the contribution from the host galaxy, propagation through the inter galactic medium and the Milky Way) should not vary periodically with the spin phase, in contrast to binary shrouding models.  Binary shrouding could generate asymmetry in the phase dependence of $\Delta$DM, which the ULP magnetar model likewise does not predict.  Likewise, phase variation of FRB fluence and frequency selection would operate at the edges of the active window in shrouding models, also in contrast to a ULP magnetar model. Furthermore, the low-twist model predicts that the period of the rotator and the dynamic range of burst energies/fluences may be anti-correlated, such that ULPs are associated with a greater abundance of low-energy bursts.

Polarization position angle variation of bursts offers a potential path to detect periodicity and falsify some models. For precession models, a shorter spin period of $1-10$ s ought to be present secularly modulated by the slower precession period. In shrouding models, barring detection of a short spin period, polarization angle variations from orbit to orbit should not be correlated with the phase of the observed period. In the ULP magnetar case, the prediction for polarization angle variations is less clear. The plasma density is so low that vacuum birefringence effects may dominate if radio emission originates from the magnetosphere. This may complicate interpretation of polarization angle variation since the eigenmode switching could occur in propagation at lower altitudes for small absolute variations in plasma density. Moderate or small polarization angle variation has been observed in FRB 121102 over various time baselines ranging up to 7 months or longer \citep{Michilli2018}.

\subsubsection{Populations of FRBs}

In our model, ULPs are associated with mature magnetars which nevertheless are still relatively young $\lesssim 10^{4}$ yr.  As in other magnetar models, FRB from ULP magnetars will be preferentially localized in regions of active star formation\footnote{Though one cannot exclude magnetars formed via non-supernova channels, such as accretion-induced collapse or binary neutron star mergers, which are delayed with respect to star formation (e.g.~\citealt{Margalit+19})}, consistent with the positions of the two best-localized repeaters, FRB 121102 \citep{Tendulkar+17} and FRB 180916 \citep{Marcote2020}. If FRB 180916 is a mature magnetar, then its older age is consistent with a much dimmer persistent radio source assuming the latter (e.g. as that detected in FRB 121102) arises from a birth nebula powered by a particle wind (e.g.~\citealt{Beloborodov17,Margalit2018}). 
In the fallback accretion scenario for ULPs, their formation could preferentially be associated with weak supernovae and/or those from massive ($\gtrsim 20 M_\odot$) progenitor stars.

Based on the observed Galactic magnetar population and the proof-of-concept model presented in this work, we predict that ULP magnetars will be rare among the canonical magnetar population (subject to large uncertainties on priors associated with initial conditions in the fallback and mass loaded wind channels for ULPs). Therefore, the period distribution inferred from a population of periodic FRBs offers a way to constrain models for their periodicity.
In addition, if an anti-correlation exists between rotation period and efficiency of burst production (e.g., as predicted in the low-twist model), then FRB samples may be be biased towards the detection of bursts from ULPs. Therefore, consideration of the FRB energy distribution  (in different bands, since frequency selection could be important) could also offer a method to constrain models. 

\subsection{Ultra long period magnetars in the pulsar population}

The observability of rotation-powered pulsars, particularly at large spin periods, is believed to be governed by the global magnetosphere's ability to produce $e^+e^-$ pairs. Neutron stars beyond the death line are in the "graveyard", which they enter once their rotational spin-down losses or magnetic field decay to the point that magnetic pair production becomes untenable, quenching their observable radio emission. In our Galaxy and its halo, ${\cal{O}}(10^9)$ ``dead" neutron stars ought to exist in the graveyard.  

Yet, the observability of magnetars is not limited by the standard polar cap curvature pair death line but by formation of current systems in their magnetospheres with significant nonzero curl (``twists") and bursts associated with dissipation of magnetic helicity and motion of the crust. Indeed, many magnetars are solely discovered as high-energy transients, later confirmed with multiwavelength scrutiny. Thus, in principle, the upper limit to $P$ for an observable active (isolated) magnetar is set by formation and evolution of applied spin torques and can extend well beyond the death line (see figure \ref{fig:MonteCarloPWspin}). Interestingly, all three spin-down channels considered in this work, {\it require} high magnetar-like poloidal fields for attainment of high spin periods. It may be that the only neutron stars with ultra-long periods are magnetars, or were so earlier in their evolutionary history. Besides the main application to FRBs considered in this work, the existence of this new population has broad implications related to their formation. 

For our Galaxy, deep X-ray surveys with frequent ($\ll 10^6$ s) revisiting cadence are likely required to discover any additional ULP magnetars such as 1E 161348--5055. This owes to the fact that pointed X-ray observations typically do not linger on candidates for $10^4-10^6$ s and, for bursts, current wide-field GRB detectors may not have sufficient angular resolution to avoid source confusion in the galactic plane. Future sensitive and dedicated wide-field X-ray transient surveyor concepts such as TAP \citep{2019BAAS...51g..85C} or THESEUS  \citep{2018AdSpR..62..191A} may alleviate these issues.

{\bf Data availability} The data produced in this study will be shared on reasonable request to the authors.

{\it Acknowledgments.} PB thanks Wenbin Lu for helpful discussions.
The research of PB was funded by the Gordon and Betty Moore Foundation through Grant GBMF5076.  ZW thanks Alice Harding, Demos Kazanas, and Andrey Timokhin for helpful discussions. ZW acknowledges support by the NASA Postdoctoral Program. BDM acknowledges support from the Simons Foundation (grant \#606260).
\vspace*{-8mm}.

\bsp	
\label{lastpage}
\end{document}